\documentclass[format=acmsmall, review=false]{acmart}
\usepackage{acm-ec-24}
\usepackage{booktabs} 
\usepackage[ruled]{algorithm2e} 
\renewcommand{\algorithmcfname}{ALGORITHM}
\SetAlFnt{\small}
\SetAlCapFnt{\small}
\SetAlCapNameFnt{\small}
\SetAlCapHSkip{0pt}
\IncMargin{-\parindent}

\usepackage{multicol}
\usepackage{graphicx}
\usepackage{caption}
\usepackage{bbm}
\usepackage{empheq}

\newcommand*\widefbox[1]{\fbox{\hspace{2em}#1\hspace{2em}}}

\setcitestyle{authoryear}

\title[Cycles and collusion in congestion games under Q-learning]{Cycles and collusion in congestion games under Q-learning}

\author{Cesare Carissimo (ETH Zurich) \newline Jan Nagler (Frankfurt School) \newline Heinrich H Nax (UZH)}

\begin{abstract}
We investigate the dynamics of Q-learning in a class of generalized Braess paradox games. These games represent an important class of network routing games where the associated stage-game Nash equilibria do not constitute social optima. We provide a full convergence analysis of Q-learning with varying parameters and learning rates. A wide range of phenomena emerges, broadly either settling into Nash or cycling continuously in ways reminiscent of `Edgeworth cycles’ (i.e. jumping suddenly from Nash toward social optimum and then deteriorating gradually back to Nash). 
Our results reveal an important incentive incompatibility when thinking in terms of a meta-game being played by the designers of the individual Q-learners who set their agents' parameters. 
Indeed, Nash equilibria of the meta-game are characterized by heterogeneous parameters, and resulting outcomes achieve little to no cooperation beyond Nash. 
In conclusion, we suggest a novel perspective for thinking about regulation and collusion, and discuss the implications of our results for Bertrand oligopoly pricing games.
\end{abstract}

\begin{document}

\begin{titlepage}

\maketitle

\setcounter{tocdepth}{2} 
\tableofcontents

\end{titlepage}

\section{Motivation and summary of main contributions}

Algorithmic collusion has been a hot topic. 
Latest since 2017 when the OECD held the inaugural roundtable on ``Algorithms and Collusion'' have scientists been working actively to improve our understanding of the issue (see, for example, the \textit{OECD 2022 Handbook on Competition Policy in the Digital Age} for its recent report).

In game theory, what is commonly dubbed as algorithmic collusion are phenomena whereby populations of intelligent agents learning to play repeated and strategic multi-agent situations achieve Pareto-superior payoffs compared with the associated stage-game Nash equilibria. 
In economics, the game context that has been studied most to investigate algorithmic collusion from this viewpoint is collusive price-setting in (generalized) Bertrand oligopoly competition games, with particular attention paid to algorithms using Q-learning 
\citep{calvano2020artificial, calvano2020protecting, klein2021autonomous}. Q-learning is widely used in practice and theoretically relatively well understood. 
Bertrand models are simple and elegant, and known to produce Edgeworth cycles under repetition \citep{maskin1988theory,maskin1988theory2}.\footnote{In the basic model that produces the duopolistic Edgeworth cycle, two firms take turns undercutting each other's prices until reaching marginal-cost levels, where they make zero profits until one of the firms drastically increases its price, which is then again followed by a process of gradual alternating undercutting, etc.} Indeed, Edgeworth cycling in Bertrand oligopolies is arguably the best-known example of tacit collusion in economics, and there is growing empirical evidence of such behavior in the real world (e.g. from gasoline pricing, airline ticket pricing, etc.) and in laboratory experiments 
\citep{kruse1994bertrand}. 

The result that Q-learning algorithms can learn these kinds of Edgeworth cycles made quite a splash since \cite{klein2021autonomous}.
It is now quite well-established that Q-learning does not generally lead to Nash in various settings. 
In particular, multi-agent Q-learning under homogeneous parameters and $\epsilon$-greedy policies with decayed exploration rates ($\epsilon$) has been shown to result in collusion on focal prices with simultaneous play \cite{calvano2020artificial}, as well as to lead to Edgeworth cycling under sequential play \cite{klein2021autonomous}.
However, these results are not robust to additional reasonable assumptions concerning algorithms' knowledge of certain key elements regarding game structure such as downward-sloping demand (or with counterfactual knowledge), as \cite{asker2022artificial} has shown for the case of deterministic policies (e.g. $\epsilon$-greedy with $\epsilon=0$). 
Related, \cite{abada2023artificial} study the setup of \cite{calvano2020artificial}, and show that when exploration is made `rational'---which means picking the Nash action when exploring instead of a uniform random action---collusion is also nullified. Indeed, it has become clear that only `pure' randomness entering $Q$-learning via an $\epsilon$-greedy policy plays a large role in the emergence of coordinated behavior. 

This observation highlights the conceptual similarity of the study of algorithmic collusion in Bertrand games by Q-learning with earlier work by \cite{wunder2010classes} (and more recently by \cite{banchio2023artificial} and \cite{dolgopolov2024reinforcement}) which consider Q-learning in Prisoner's Dilemmas. A key result of that literature has been that maintaining a constant exploration rate ($0 < \epsilon \leq 0.1$) can lead to `collaboration' and that small values of the exploration rate lead more frequently to Pareto improvements. Recent results by \cite{dolgopolov2024reinforcement} underpin these results theoretically.

In this paper, we investigate Q-learning in another important class of games, namely in many-player games generalizing the game underlying the Braess Paradox \cite{braess1968uber}. 
In our own prior work on this game, we found similar results to be robust to many players ($n=100$), with small (non-decaying) exploration rates resulting in cycles and Pareto improvements \cite{carissimo2024counter}.
In this paper, going beyond \cite{carissimo2024counter} whose basic set-up we extend in several key aspects, we 1) characterize the cyclical behavior of continual $Q$-learning as dependent on learning rates, 2) analyze the effects of receiving semi-bandit feedback, 3) explain the emergence of cycles as dependent on delays, 4) formulate the parameter-picking meta-game, and 5) test the existence of symmetric pure-strategy Nash equilibria in the parameter-picking meta-game.

We therefore fit into the emergent literature on algorithmic collusion as follows. Following \cite{wunder2010classes, banchio2023artificial, dolgopolov2024reinforcement}, we consider non-decaying exploration rates. This continual learning approach we believe is relevant in situations where it is realistic to assume that algorithm designers seek robustness to an underlying payoff structure that might change in time and/or where algorithms partially observe the environment. Going beyond what prior work has done, we determine the effects that learning rates have on cyclical dynamics and collusion in the Braess Paradox. Then, we test whether homogeneous parameters are incentive compatible in the parameter-picking meta-game. We are not aware of prior work doing this, as all the aforementioned papers assume identical algorithms with identical parameters, and their reported ablation studies vary parameters identically for all players.
Indeed, it is this aspect of our study that reveals the key insight of our paper:
there is an important incentive incompatibility when thinking in terms of a meta-game being played by the designers of the individual Q-learners who set their agents' parameters. 
Indeed, Nash equilibria of the meta-game are characterized by heterogeneous parameters, and resulting outcomes achieve little to no cooperation beyond Nash. By contrast, the homogeneous parameter combinations that result in outcomes with high social welfare and cycles do not constitute Nash equilibria.
In conclusion, we therefore suggest a novel perspective for thinking about regulation and collusion, and discuss the implications of our results for Bertrand oligopoly pricing games.

\section{Model in a nutshell}

Our game is played as follows.
Each player in a population repeatedly picks routes in a network to get from one origin to one target with the aim of minimizing their travel cost. 
The network is simplified 
following
Braess so that each player has the same three choices of paths, two of which (``up'' and ``down'') are socially optimal when chosen equally by the population, while the third path (``cross'') is individually faster given any vector of others' choices but socially sub-optimal when it leads to overuse of either up or down. As such, all players picking ``cross'' is the Nash Equilibrium, and characteristic of Braess-like games it is Pareto inefficient. The paradox is revealed by removing ``cross,'' which reduces the network's capacity but results in a Pareto efficient Nash equilibrium. 

It is relevant in our digital age to study shared resource games played by learning agents. 
Imagine, for example, that independent internet routers are controlled by algorithms as they choose among paths to transmit their packages on.
We may also imagine that this set-up describes the routing decisions of completely decentralized autonomous vehicles' on behalf of, for example, commuters traveling daily from one city to another and choosing between two sets of minor roads or a highway overpass. 
In 
those scenarios, the routers' or drivers' decisions 
may critically
determine the costs incurred by each action: paths may congest when chosen by many, and travel 
may be 
quick on paths chosen by few. Additionally, it is reasonable to assume that a `model free' approach to training neural networks with reinforcement learning is used, because the internet and traffic are complex systems which are challenging to model \textit{apriori}.
$Q$-learning, as the simplest form of reinforcement learning capable of handling complex environments (Markov Decision Processes), gives us a window of experiments to study where learning converges when sharing resources, and with which dynamics. 
In games like the Braess Paradox with a Pareto inferior Nash equilibrium, it allows us to evaluate the Price of Anarchy for a more general class of agents which is \textit{explicitly}\footnote{As opposed to best response and/or regret which assume behaviour with a theoretical guarantee.} doing basic machine learning.

Our set-up uses $Q$-learning agents with an $\epsilon$-greedy policy. Our game is stateless such that agents only store the value of each action (``up'', ``down'', ``cross''), and have no state information which can reveal the actions of other players. Agents only receive feedback for their actions and those of other players through their rewards (travel time). We adopt a continual learning framework by keeping a constant and positive exploration rate throughout the entire learning process for all agents.

One question we address in this paper is whether a population of Q-learners will learn to play this Nash equilibrium or whether it will achieve Pareto superior outcomes. Indeed, we find that Q-learning can lead to Edgeworth-like cycling. Importantly, other than in some prior work (e.g. \cite{klein2021autonomous}) these cycles are not the result of convergence to Markov equilibria, but instead are produced by cycles in the learning patterns themselves.
Here, 
we analyze 
the effects of 
different learning rates for 
different information models. 
Specifically, we study how the learning rates of individual $Q$-learners influence the outcomes for a population. 
%

Individual learners learn in two ways; `self-learn' based on own feedback (learning rate $\alpha$ for bandit learning), and `other-learn' based on others' feedback (a new parameter $\beta$). We find that populations of Q-learners with fast individual self-learning and slow other-learning produce the conditions that foster cycling with average payoffs above stage-game Nash
equilibrium, 
while slow individual self-learning and fast other-learning result in low payoffs and Nash equilibrium.  
Individually, however, having a slower self-learning rate and a faster other-learning rate constitutes a unilateral advantage. 
In the remainder of this paper, we shall make these results explicit, and discuss their consequences for optimal system design and regulation. 

The paper is organized as follows. In the next section, we set up the game and \textit{continual} Q-learning formally. 
In Section 4, we fully characterize how and when cyclical behavior emerges for homogeneous populations of Q-learners. In 
Section 5, we analyze the incentive compatibility the homogeneous case. Finally, in section 6, we discuss the consequences of 
our findings 
for regulation and design of 
congestion games and for games with 
similar 
payoff structures like Bertrand oligopolies. 

\section{Methods}

In this section we describe the underlying game and our implementation of $Q$-learning.
The underlying game is a general congestion model with the cost structure of the Braess Paradox. 
Our implementation of $Q$-learning is \emph{continual}, and we shall describe the
numerical experiments 
and
observables
that will assessed in our results section. 

\subsection{Congestion Games and the Braess Paradox}

\begin{figure}
  \begin{minipage}[b]{.45\linewidth}
    \centering
    \includegraphics[width=0.7\linewidth]{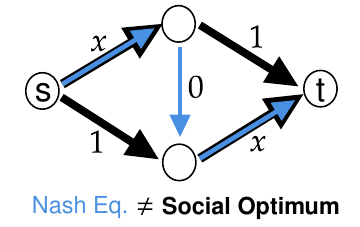}
    \captionof{figure}{The Braess Paradox network where three paths are possible: up, down and cross. Splitting 50\% of the population up, and the other 50\% down is socially optimal, but the Nash equilibrium has all agents picking cross. Numbers represent link costs, where "x" is the fraction of total agents $f(a)/n$.} \label{fig:network}
  \end{minipage}\hfill
  \begin{minipage}[b]{.45\linewidth}
    \centering
    \begin{tabular}{ |c|c| }
     $a$ & $u_a$ \\
     \hline
     up & $1 + \frac{f(\text{up}) + f(\text{cross})}{n}$ \\[0.2cm]
     cross & $\frac{f(\text{up}) + 2f(\text{cross})+f(\text{down})}{n}$ \\[0.2cm]
     down & $1 + \frac{f(\text{down}) + f(\text{cross})}{n}$ \\
    \end{tabular}
    \vspace{0.5cm}
    \captionof{table}{Actions $a$ and their costs $u_a$ where $f(a)$ counts the number of agents that select the action.}\label{tab:payoffs}
  \end{minipage}
\end{figure}

General congestion games \citep{milchtaich1996congestion} are non-cooperative games where $n$ players share a set $\mathcal{A}$ of strategies representing routes, and the payoffs that are associated for each player with every route depend on the number of other players that travel them as described by a function $u_a(f(a))$, where $f(a)$ represents the number of players playing strategy $a\in\mathcal{A}$, and the payoff function $u_a$ is monotonically decreasing in $f(a)$. In this paper we focus on a specific congestion game, the generalized Braess Paradox, which has three actions, $\mathcal{A}=\{0, 1, 2\}=\{\text{up}, \text{down}, \text{cross}\}$, representing two symmetric regular routes and a high-capacity route with payoff functions as described in Table \ref{tab:payoffs} \cite{carissimo2024counter}. We call the vector specifying an action for each player the action profile $\textbf{a} = (a_1, \dots, a_n)$.

Congestion games, in general, have unique pure-strategy Nash equilibria \citep{milchtaich1996congestion}, part of a larger class of potential games \citep{monderer1996potential}. The Nash 
equilibrium of the generalized Braess Paradox in \autoref{fig:network} implies
all agents to play \textit{cross}, meaning,
$f(\text{up})=f(\text{down})=0$ and $f(\text{cross})=n$.
Then,
all agents receive a travel cost of $2$. 
Because 
both paths \textit{up}, and \textit{down}, also have travel times of $2$, 
in the Nash 
equilibrium all players are effectively indifferent between the actions. 
The Nash 
equilibrium is Pareto inefficient because
the socially optimal action profile
requires 
half of the agents to 
choose \textit{up}, 
and the other half 
to pick \textit{down}, $f(\text{up})=f(\text{down})=n/2$.
For this action profile all agents obtain a travel cost of $1.5$. 


\subsection{Q-learning in the Braess Paradox}

To study the behaviour of $Q$-learning agents in the Braess Paradox network we initialize a population of $n$ $Q$-learners \cite{watkins1992q, sutton2018reinforcement}. Each $Q$-learner is given a $q$-function, $Q:\mathcal{A}\rightarrow \mathbb{R}$ which maps actions to real numbers. Given a small action set $\mathcal{A}=3$ the $Q$-function is conveniently represented as a length $3$ vector which we here-on-out refer to as a $q$-table for each agent. 

The $Q$-learners repeatedly play the Braess Paradox congestion game $T$ times. At the start of the game, all agents use their policies $\pi_i$ to select an action $a_i$, $\mathbf{a} = (\pi_1, \cdots, \pi_N) = (a_1, \cdots, a_N)$. The agents use an $\epsilon$-greedy policy function $\pi$, as has also been used in the related literature on pricing games \citep{calvano2020artificial, klein2021autonomous}. 
The 
policy picks 
an action uniformly at random with probability $\epsilon$,
and otherwise, with probability $1-\epsilon$, 
$arg\max_a Q(a)$.
Subsequently, 
the agents 
receive the reward 
for
their action 
$\mathbf{r} = (r_1, \cdots, r_n)$, according to 
the reward function $\mathcal{R}:\mathcal{A}^{n} \rightarrow [1, 2]$,
which simply maps costs (journey time) to rewards based on 
the costs 
as specified in Table \ref{tab:payoffs}.
Then, the 
$q$-value $Q_i(a_i)$ is gradually updated,
\begin{equation}\label{eq:bellman}
Q(a_i) \leftarrow Q(a_i) + \alpha(r - Q(a_i)),
\end{equation}
where the only parameter
$\alpha$ controls
the learning rate. 

Applications for $Q$-learning include 
pricing games \cite{calvano2020artificial, klein2021autonomous} 
and the Prisoners Dilemma \cite{dolgopolov2024reinforcement}, 
where $Q$-learners receive \textit{bandit feedback},
which means that $Q$-learners only learn from the actions they take. 
Here, 
we show that 
the standard approach of
bandit feedback creates oscillatory delay dynamics,
which we discuss in \autoref{sec:cycles}. 
In contrast to previous research, we also study a \textit{full monitoring} case where agents receive feedback for all actions including 
even
the ones they did 
{\em not} take. The updates to the $q$-values of agents for the actions they did not take is parametrized by the additional 
parameter $\beta$. 

\paragraph{($\beta$) The Monitoring Feedback Delay Rate} 
For an agent $i$, the reward 
for her 
picked action
$a_i$ 
is received immediately, but the feedback for the actions that were not picked by the agent, which we denote as $a_{-i}$, 
ought to be rewarded delayed.
Concretely,  for agent $i$
the actions $a_{-i}$ are updated depending on 
$\beta$ as follows:

\begin{equation}\label{eq:monitoring}
Q(a_{-i}) \leftarrow Q(a_{-i}) + \beta(r - Q(a_{-i}))    
\end{equation}

Therefore, 
the additional 
parameter $\beta$ 
encodes 
the delay rate of the monitoring feedback.
Here, 
we require 
that $\beta$ is always smaller than $\alpha$, such that the $Q$-learners always react slower to monitoring feedback than to the
immediate rewards \footnote{\textit{Note} that
the fundamental sequence of $q$-table updates is unaltered 
by our approach. 
We do {\em not}
explicitly introduce any additional delay. In particular, all
updates of $q$-tables are synchronous. For agent $i$, all actions are updated at every time-step: 
action $a_i$ with \autoref{eq:bellman}, and actions $a_{-i}$ with \autoref{eq:monitoring}.}.

\subsection{Parameter Picking for the Repeated Game}

There are two parameters that $Q$-learning algorithm designers pick have a strong influence on the outcomes of the repeated route-picking game: $\alpha$ and $\epsilon$. Think of the whole situation as a two-stage game. In stage 1, a one-shot simultaneous-move meta-game takes place in which the parameters and learning rates of each Q-learner in the population are picked individually by competitive designers. In stage 2, the repeated route-picking game that we studied throughout the paper takes place, in which each agent is characterized by its individual parameters and learning rate. Average payoffs of the stage-2 game as earned by the individual Q-learners represent the payoffs of the meta-game in stage 1, which is the game we study in this section. 

To determine the effects that parameter choices have on the outcomes of the repeated game, we denote the time-averaged travel time for player $j$ as:
\begin{equation}
	\langle C_j \rangle = \frac{1}{T} \sum_{t=1}^T r_{j,t}.
\end{equation}
The \textit{average travel time} of the system is: 
\begin{equation}
	C_t = -W_t = \frac{1}{n}\sum_{i=1}^n r_{i,t}.
\end{equation}
which measures the instantaneous system cost, or negative social welfare, at any time $t$. The average travel time of the system changes in time due to the dependence of the actions taken by the $Q$-learners. We also define the the \textit{time-averaged travel time}:
\begin{equation}
	\langle C \rangle = \frac{1}{T} \sum_{t=1}^T C_t.
\end{equation}
which reflects the overall system performance over the entire learning process. In \autoref{sec:cycles} we will analyse the case where parameters are picked homogeneously for the population of $Q$-learners. Then in \autoref{sec:incentives} we extend our analysis to some strategic heterogeneous cases.

\subsection{Discussion of Learning Rates}

In section we will focus on the choice of the learning rate $\alpha$, the main control parameter for the learning process. Given an action $a$ and a reward $r$ for taking action $a$,  at a time $t$, the learning rate $\alpha$ determines the extent to which $r$ is used to update the $q$-value $Q(a)$ using Equation \ref{eq:bellman}. We assume that 
$\alpha$ is a constant parameter 
over
time, but may vary heterogeneously between agents. The learning rate can be seen as encoding how much an agent believes the last 
received
reward to reflect the 
reward structure 
that fundamentally underlies the game.
Thus, the learning rate guides the convergence of an 
agent's beliefs 
$Q(\cdot)$ 
to the games' 
reward structure. 

Picking an $\alpha$ implies 
making an assumption 
on 
the informativeness of rewards. For example, picking $\alpha=1$ means that given an action $a$, a reward $r$, and a $q$-value $Q(a)$, the $q$-value will be updated directly to $r$ loosing all memory of the previous $q$-value: $Q(a)\leftarrow Q(a) - Q(a) + r = r$. On the other hand, picking $\alpha=0$ means the $q$-value will never be updated. For values of $0<\alpha<1$, the $q$-values start to have \textit{memory} which traces a history of rewards like an exponential moving average: 

\begin{empheq}[box=\widefbox]{align*}
    & q-\textbf{values retain memory with } \alpha-\textbf{exponential decay} \\
    Q_t(a) = \: & Q_{t-1}(a) + \alpha(r - Q_{t-1}(a)) \\
    = \: & Q_{t-2}(a) + \alpha(r_{t-2} - Q_{t-2}(a)) + \alpha(r_{t-1} - Q_{t-2}(a) + \alpha(r_{t-2} - Q_{t-2}(a))) \\
    = \: & (1 - \alpha)Q_{t-2}(a) - (\alpha + \alpha^2)Q_{t-2}(a) + \alpha (r_{t-2} + r_{t-1} - \alpha r_{t-2}).
\end{empheq}

Suppose our exponential moving average has converged to the underlying value $x$, and that the underlying value now changes and becomes $y$: it will take on average $1/\alpha$ updates for an exponential moving average to
approach 
$y$. An exponential moving average is well suited for a non-stationary target which changes over time as past observations contribute exponentially less to the average as a function of time. Hence, 
$\alpha$ can also be interpreted as the expected 
number of observations that an agent needs 
for learning 
the underlying reward structure of the game. 

Having covered the meaning of the learning rate $\alpha$, let us take the perspective of a single agent 
in a population of learning agents: a router on a network of routers, or 
be it an automated vehicle in a traffic network. 
For these cases, 
we wish to ask the following question from the perspective of a single $Q$-learner: given that all other agents pick their learning rates, which learning rate should I pick? We have explained that picking $\alpha$ is a statement on the expected number of observations $1/\alpha$ which
is 
needed to trust that the $q$-values have converged to the true underlying value. So in the case of $Q$-learners in a congestion game, we must identify what the true underlying value is. While agents receive feedback as a reward/cost for their independent travel times, these costs depend on the actions of all other agents. Given that all other agents are also using $Q$-learning, their actions may be both stochastic and non-stationary. 
After all, in a stochastic (largely uncertain) 
environment 
we may prefer to choose 
a rather small $\alpha$, whereas if the 
environment is rather predictable, 
we may afford to pick a comparably larger $\alpha$. In other words, in a stochastic environment we trust each individual observation less, thereby expecting 
to need many observations before
we have learned 
the true underlying distribution which generates the observations. Picking a larger $\alpha$ allows us to converge faster while sampling 
the observations, but it may be less suited 
in 
a strongly fluctuating 
environment. 

In a stationary environment, all conceivable average observables (expected values) characterizing the environment are time-invariant. 
In a non-stationary environment, there may be a trend that renders averages 
drifting in some direction, or other effects that destroy the time invariance of
some average quantities.

For such a non-stationary environment, we may have to assign $\alpha$
a different role.
Assume, our non-stationary environment has an underlying target which changes
over time, like a trend. Yet, 
the 
target may also be stochastic. 
If the target changes 
over time,
then we may want 
to more rapidly 
forget about the past and update our $q$-values to the present. 
This means, however, that 
an agent 
may actually benefit from a higher $\alpha$ which weights that past less. Therefore, in a congestion game like the Braess Paradox where rewards are both stochastic due to an intrinsic randomness of player action policies, and non-stationary due to the cyclical behaviour induced by delays in $Q$-learning with imperfect monitoring, there is 
quantifiable 
tension between the stochastic and non-stationary nature 
of the rewards. 

\subsection{Measuring Cycles}












In this section we will define the metrics that we use to evaluate the effects that our parameters have on the period and asymmetry of the cycles that emerge during $Q$-learning.

The \textit{period} reflects the average number of steps it takes to complete a full cycle. To calculate the period we first take the time-averaged travel time $\langle C \rangle$, and count the number of times that the 
time series $\{C_t\}_{t\in |T|} = \{C_1, C_2, \dots, C_T\}$ crosses 
$\langle C \rangle$
from above, excluding all steps with an absolute difference smaller than $3\sigma$, where $\sigma$ is the standard deviation of the differences of $\{C_t\}_{t\in |T|}$ between subsequent timesteps, $\{\delta_t\}_{t\in |T|} = \{\delta_1, \delta_2, \dots, \delta_T\}$.

\begin{equation}
    M = \mathbbm{1} \big\{ (C_{t} > \langle C \rangle) \: \cap \: (C_{t+1} < \langle C \rangle) \: \cap \:  (|\delta_t|>3\sigma)\big\}
\end{equation}

Thus, $M$ estimates the number of cycles in $\{C_t\}_{t\in |T|}$. We calculate the frequency of these cycles $\omega=\frac{M}{T}$. Finally, we determine the period of these cycles: 

\begin{equation}\label{eq:period}
    L = \frac{1}{\omega} = \frac{T}{M}.    
\end{equation}

The ``\textit{edgeworthiness}" reflects the relative frequency of price increases in the time series $\{\delta_t\}_{t\in |T|}$, to capture whether or not the time series contains many gradual increases and few sharp decreases. Our measure of edgeworthiness is a measure of asymmetry. Asymmetry is frequently searched for to detect Edgeworth cycles in real-world data \cite{holt2021detecting}. To measure this asymmetry we count the number of price increases in $\{\delta_t\}_{t\in |T|}$, and divide it by the total length of $\{\delta_t\}_{t\in |T|}$, $T-1$:

\begin{equation}
    F = P(\delta_t >= 0) = \frac{|\big\{ \mathbbm{1}\{\delta_t >= 0\}_{t\in |T|} \big\}|}{T-1}
\end{equation}

\subsection{
Technical and conceptual contributions} 

The recent literature on $Q$-learning in games \cite{calvano2020artificial, klein2021autonomous} has simulated the convergence properties of $Q$-learning algorithms in Bertrand duopolies, and they achieved this by 
keeping 
the learning rate $\alpha$ 
and discount factor $\gamma$ constant, and then by decaying the exploration rates $\epsilon$.
In 
doing so, the $Q$-learners start playing the repeated duopoly with a very high exploration rate, close 
to
$1$, which implies a near-random behaviour with an $\epsilon$-greedy policy. While $\epsilon$ is exponentially decayed, the $Q$-learners actions become increasingly deterministic, determined solely by the current state they perceive. By the end of the simulation, the algorithms have converged on deterministic policies. In the case of \cite{calvano2020artificial}, these policies 
generated
prices which were higher than the competitive equilibrium, while 
\cite{klein2021autonomous} additionally showed that in a sequential duopoly these policies converged 
to
Edgeworth cycles. Therefore, the authors achieved this \textit{implicit collusion} with fully trained algorithms which converged to supra-competitive behaviour. It is clear from their setup that the focus was on the behaviour of trained algorithms, and not on the behaviour of algorithms during training.

Our proposed framework differs, 
because we study the online, \emph{continual} learning of a system of $Q$-learners which have not converged, nor 
they actually will 
converge. We achieve this with 
a constant exploration rate $\epsilon$ rather than decaying it. We do so, following 
the steps of recent literature \cite{carissimo2024counter}. In studying the online behaviour of the $Q$-learning algorithms, we are interested in the dynamics that emerge during the learning process: how numerous simultaneously learning algorithms lead to coordinated behaviour, as we are assuming that the algorithms keep learning throughout the simulation, and that they do not ultimately converge to any deterministic policy. 
Notably, we argue that this is justified for studying 
the Braess Paradox, a system in which many $Q$-learning agents which create a highly stochastic dynamics. In the real-world networks where the Braess Paradox may appear, the network payoff structures may 
be subject to non-stationarity, which requires learning agents to keep adapting, and therefore undergo \textit{continual}-learning.

Conceptually, we add another dimension to the analysis relevant also to the aforementioned literature. While the literature so far has focused on investigating the strategic coordination that may occur when algorithms interact in a game, we also add a layer of strategic analysis as applied to the layer of algorithm design. In other words, the parameter choices of the algorithm designers have previously not been analyzed in terms of the incentives given the parameter choices of the other designers, and the literature focuses on homogeneous parameters for all algorithms. This paper extends the analysis to determine whether the homogeneous parameter case, treated most frequently in the literature, is an equilibrium of a `meta-game' of parameter picking that we introduce.

\section{System Performance and Edgeworth Cycles}\label{sec:cycles}

For all experiments,
we set $\alpha$, $\beta$ and $\epsilon$ for $n=100$ $Q$-learners in the Braess Paradox network, and simulate their learning for $T=100000$ learning iterations. $Q$-values are initialized randomly or at the Nash Equilibrium payoffs. Each parameter combination is repeated $40$ times to account for the randomness introduced from $Q$-value initialization, and from exploration during learning. Throughout each run all parameters are kept constant.

We run experiments in the continual $\epsilon$-greedy $Q$-learning framework (in contrast to previous literature on pricing games which used decayed exploration rates \cite{calvano2020artificial, klein2021autonomous}). Previous literature found that low values of $\epsilon$, and at best an $\epsilon=0.01$ led to the most coordinated behaviour in the Braess Paradox \cite{carissimo2024counter}. Thus, we pick $\epsilon=0.01$ to vary the learning rates in a regime with a demonstrated potential for coordination.


\subsection{
Edgeworth Cycles 
and 
Delay Machanisms}\label{sec:cycles-delay}

\begin{figure}[]
    \centering
    \begin{multicols}{2}
    \includegraphics[width=\linewidth]{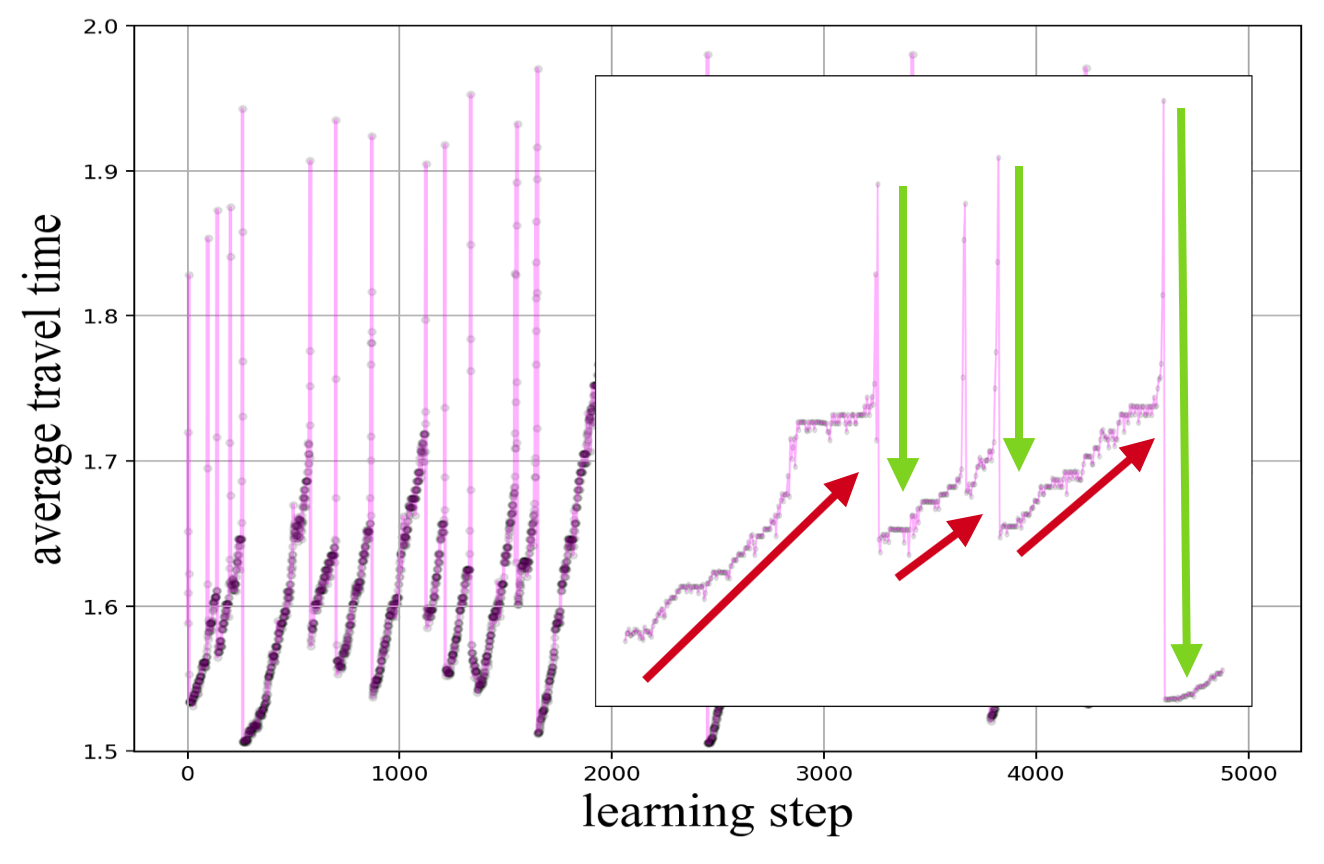}\par
    \includegraphics[width=\linewidth]{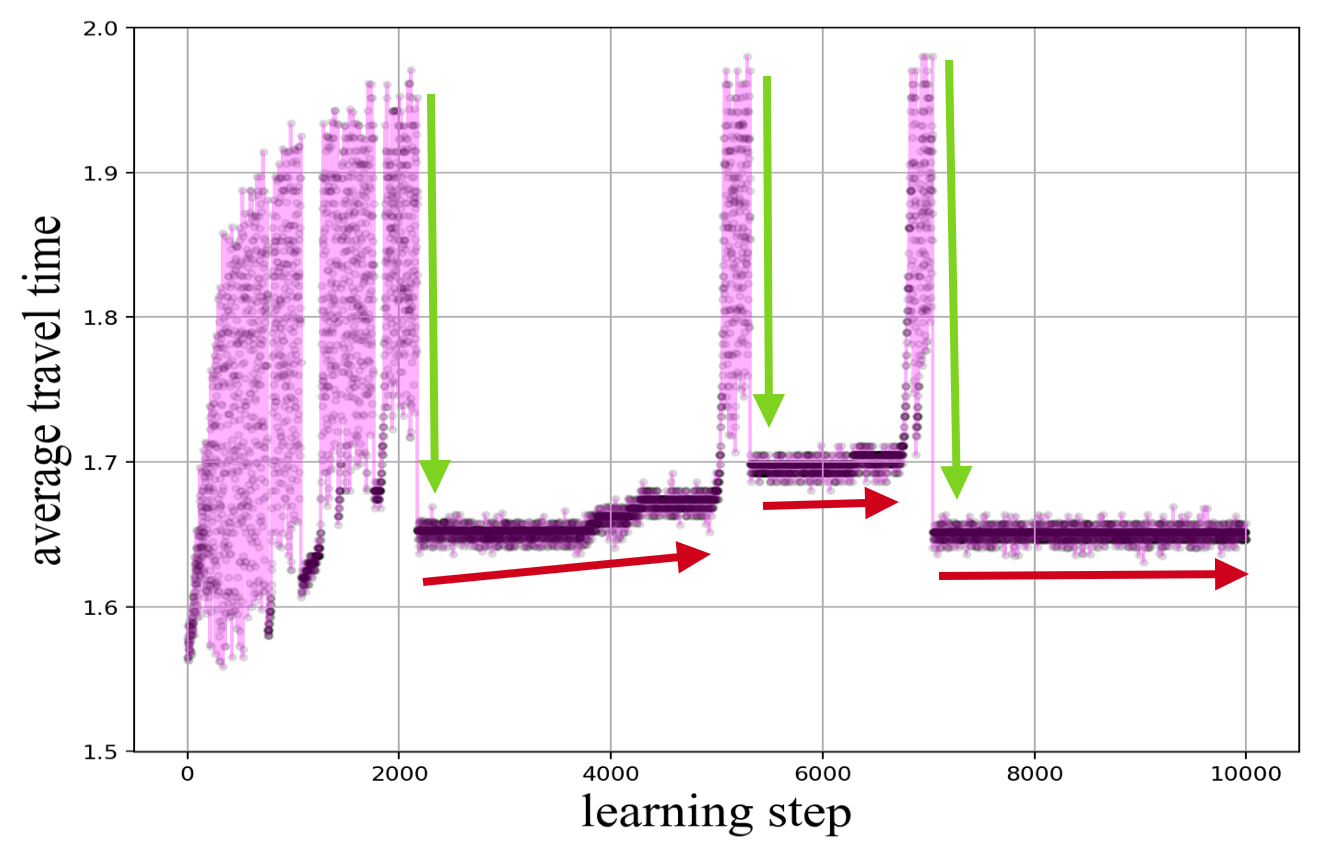}\par
    \end{multicols}
    \caption{100 Q-learners in Braess Paradox $(\epsilon=0.01, \beta=0)$ display the characteristic behaviour of Edgeworth Cycles. Left: $\alpha=0.7$, a relatively fast learning rate which leads to cycles with short periods $L$, and high asymmetry $F$: steep regions approaching the one-shot Nash. Right: $\alpha=0.01$, a relatively slow learning rate which leads to cycles with long periods $L$, and less asymmetry $F$: flat regions approaching the one-shot Nash.}
    \label{fig:oscillations}
\end{figure}

In \autoref{fig:oscillations} we show the characteristic 
behaviour of Edgeworth cycles replicated by $Q$-learners in the Braess Paradox. In our system, these cycles are caused by delays. 
It is known that delays can cause cycles and chaotic behaviour in physiology \cite{mackey_oscillation_1977, foss_multistability_1996}, biology  \cite{macdonald_biological_2008}, physics \cite{ikeda_high-dimensional_1987, simonet_transition_1995, giacomelli_relationship_1996, bestehorn_order_2000} as well as in economics \cite{mackey_commodity_1989}. 
In those examples,  
delays have been expressed as an integral part of the 
dynamical system in
terms of differential equations which
explicitly depend on the past, 
$\dot{x} = F(x(t), x(t - \tau))$, such that the system state 
$x$ depends on a $\tau$-delayed version of itself \cite{manffra2002properties}.
In contrast, 
the delay mechanisms and observed time scales 
in our study of Edgeworth cycles in congestion games
are not a direct consequence from an imposed 
delay, nor we envision a straightforward way to formulate the system
as a dynamical system -- they are emergent stochastic phenomena.

\begin{figure}
    \centering
    \begin{multicols}{3}
        \includegraphics[width=\linewidth]{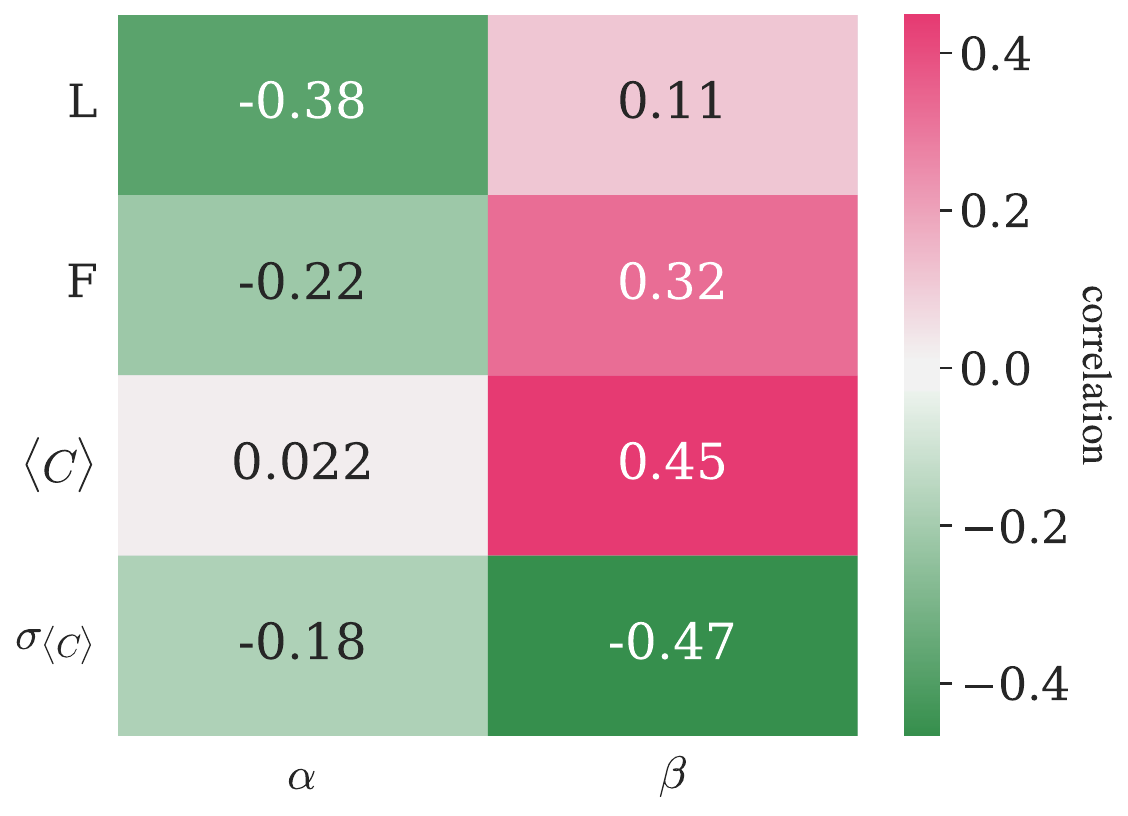}\par
        \includegraphics[width=\linewidth]{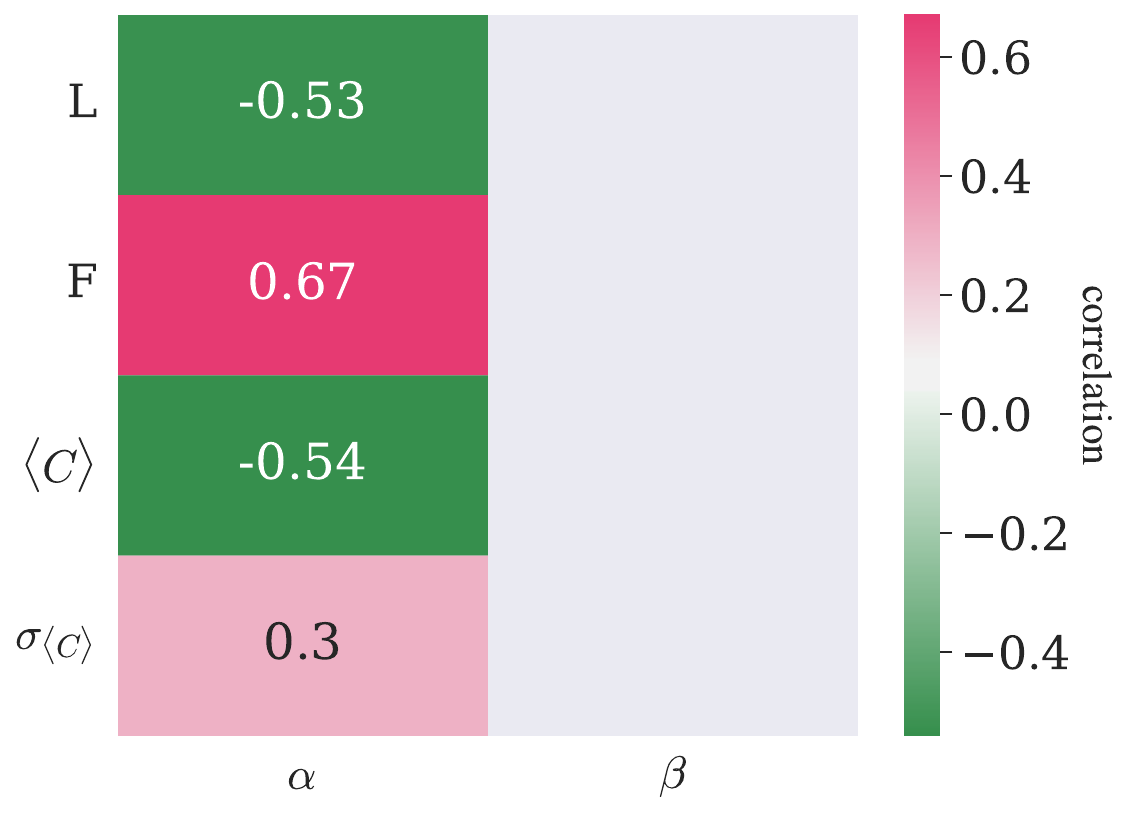}\par
        \includegraphics[width=\linewidth]{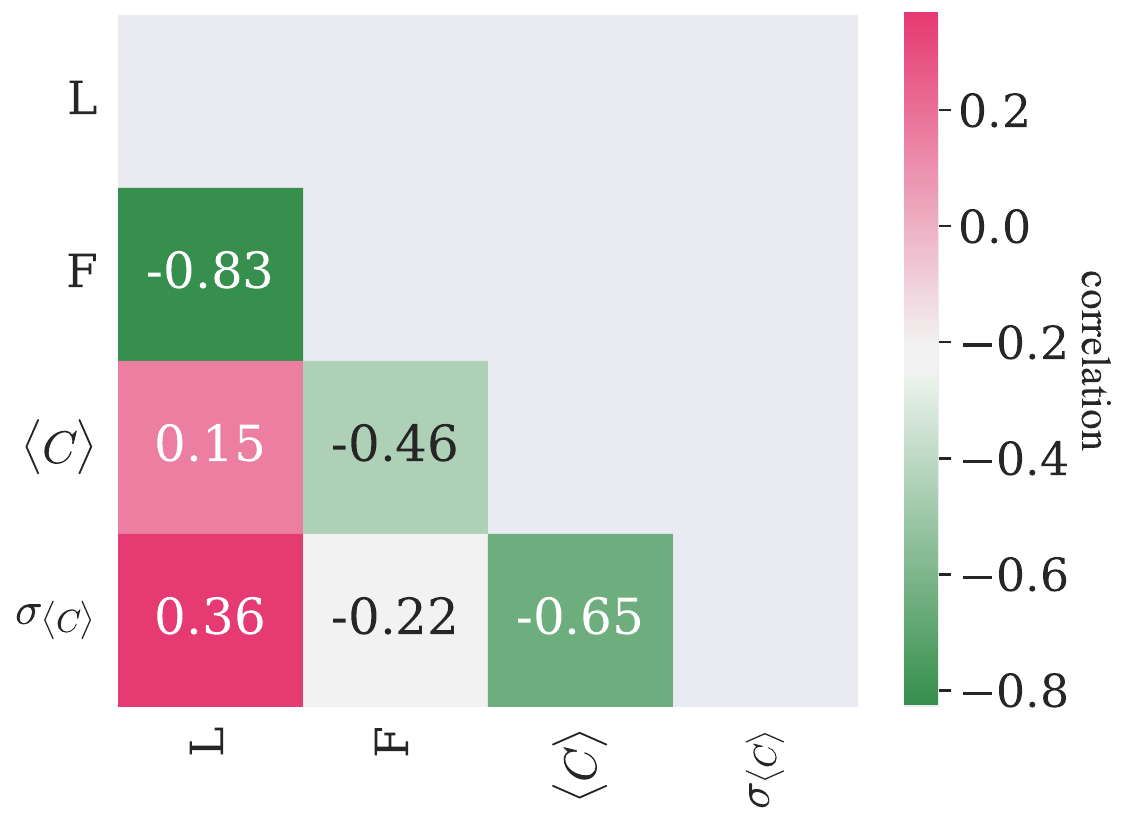}\par
    \end{multicols}
    \caption{Correlation matrices between dependent and independent variables. Left: Correlations of $\alpha$, $\beta$ with $L$, $F$, $\langle C \rangle$, and $\sigma{\langle C \rangle}$. Middle: Correlations of $\alpha$, when $\beta=0$, with $L$, $F$, $\langle C \rangle$, and $\sigma{\langle C \rangle}$. Right: Correlations of with $L$, $F$, $\langle C \rangle$, and $\sigma{\langle C \rangle}$ with themselves. The variable $\sigma_{\langle C \rangle}$ is defined as the standard deviation of $\langle C \rangle$, $\sigma_{\langle C \rangle} = \sqrt{\langle C \rangle^2 - \langle C^2 \rangle}$.}
    \label{fig:correlation}
\end{figure}

For our $Q$-learning system, each player undergoes a 
\textit{learning loop}: 
start 
with three $q$-values in 
her $q$-table $\mathbf{Q}_t=(Q_{t}(1), Q_{t}(2), Q_{t}(3))$, 
which are 
dynamic time-dependent variables indexed
for 
time $t$ \cite{wunder2010classes}. The learning loop then applies an $\epsilon$-greedy policy $\pi$ which maps the $q$-table to a single action $a_t$, followed by a feedback $r_t$ and an update to the corresponding $q$-value, $Q_{t}(a)$, using \autoref{eq:bellman} to obtain $Q_{t+1}(a)$. Repetition of the learning loop naturally creates a time-dependent dynamics for the time series of actions $\{a_t\}_{t\in |T|}$ , where actions in the future, $a_{t+1}$ depend on $q$-values which were last updated at time $t$.

\begin{empheq}[box=\widefbox]{align*}
    & Q-\textbf{learning loop creates delay and hysteresis:} \\
  a_t &= \pi[(Q_t(1), \: Q_t(2), \: Q_t(3))] = 1 & \text{(apply policy)} \\
    r_t &= \mathcal{R}((a_t)) & \text{(get reward)} \\
    Q_{t+1}(1) &\leftarrow Q_t(1) + \alpha(r_t - Q_t(1)) & \text{(update } q-\text{value)}\\
    a_{t+1} &= \pi[(Q_{t+1}(1), \: Q_t(2), \: Q_t(3))] & (a_t \text{ depends on }t \text{ and } t+1)
\end{empheq}

Therefore, past time-steps $t$ can influence future
time steps 
$t+\tau$,
even for arbitrary time horizons $\tau$. We identify that hysteresis in $Q$-learning is modulated by two endogenous parameters $\alpha$ and $\beta$, and two exogenous features of the repeated game played by $Q$-learners, stochasticity and non-stationarity as perceived by individual $Q$-learners. As the $Q$-learners try to learn the values of the rewards for actions, they will be updating their $q$-values from reward feedback, which is stochastic and non-stationary as it depends on the actions of all other $Q$-learners.

\subsubsection{Stochasticity} The repeated game environment is exogenously stochastic, driven by the exploration of $Q$-learners with $\epsilon$-greedy policies. 
At the same time, the interactions of the $Q$-learners,
which update their $q$-values on the basis of noisy reward signals, leads to endogenous dynamics which are stochastic \cite{carissimo2024counter}. For this reason, the $Q$-learners need many observations of reward for actions to determine the "true" underlying value of actions,
which contributes 
to a subtle delay mechanism imposed on the collective system's dynamics.

\subsubsection{Non-stationarity} The repeated game environment is also non-stationary from the perspective of individual $Q$-learners. 
The non-stationarity arises in the system as the "true" 
underlying value of actions changes in time. In other words, if we think of the "true" underlying value as a target, the target is moving in time from the perspective of individual $Q$-learners. In \autoref{fig:oscillations} we show how this can arise when the average travel time undergoes Edgeworth cycles. In fact, we can 
identify 
two different 
manifestations
of non-stationarity. 
In the left panel 
in \autoref{fig:oscillations},
we see a target which is gradually changing, with a high asymmetry and small period cycling. In the right panel, 
we observe 
a target which stays nearly constant for long periods, 
but the target changes more 
abruptly 
in a step-wise manner. 
The observed non-stationary behaviour results from $Q$-learning based 
on
an exponential moving average $q$-value updating.
This type of memory forgetting 
leads to the collective adaption of the $Q$-learners
to moving targets, 
which also contributes to the emergence of delays.

\subsubsection{Learning rate $\alpha$} 
The learning rate $\alpha$ controls 
the degree of how aggressively 
$q$-values are updated toward 
the most recent observation.

That means, 
$\alpha$
encodes the amount of observations required to shift the $q$-value to reflect the new underlying mean of the reward for an action. 
In particular, 
a large $\alpha$ will rapidly adopt the latest observation. This can induce rapid changes to $q$-values, and therefore rapid changes to actions. As such, $\alpha$ can influence the stochastic behaviour of the system: many agents rapidly changing their actions create noise. Conversely, a small $\alpha$ leads to slow change of $q$-values, which can reduce the stochastic behaviour of the system. We show in \autoref{fig:oscillations} that $\alpha$ also influences the non-stationarity of the system, where a high $\alpha=0.7$ leads to small period cycling with higher asymmetry, and a low $\alpha=0.01$ leads to large period cycling and less asymmetry.

\subsubsection{Learning rate $\beta$} 
The learning rate $\beta$ has a 
profound 
effect on the hysteresis of the system. This is because it is capable of completely removing any delay in $q$-value updates, where a $\beta=\alpha$ leads to all $q$-values being updated at the same time-step, irrespective of the action taken. This influence, as we will see in the next results, is capable in the extreme of completely 
removing  
the Edgeworth cycles, 
by suppression 
of both the stochastic behaviour and non-stationarity.

\subsection{How Cycles Influence Performance}

\begin{figure}
    \centering
    \begin{multicols}{2}
        \includegraphics[width=\linewidth]{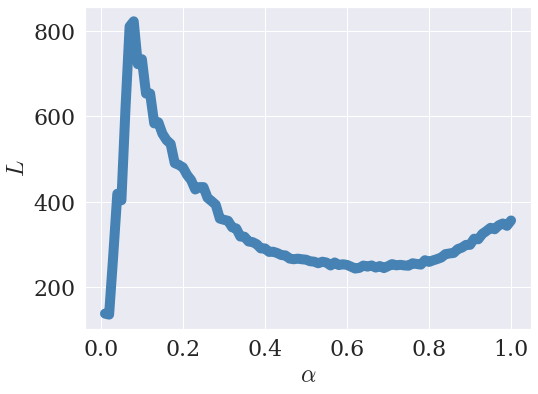}\par
        \includegraphics[width=\linewidth]{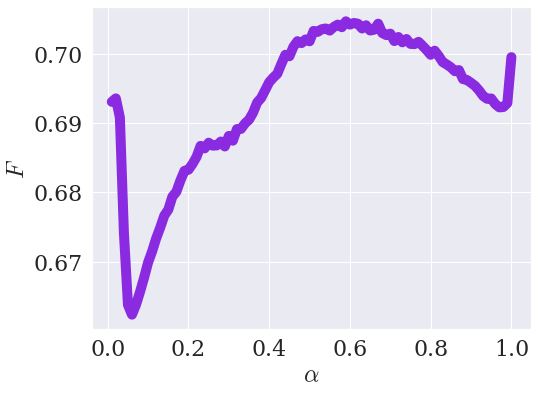}\par
    \end{multicols}
    \begin{multicols}{2}
        \includegraphics[width=\linewidth]{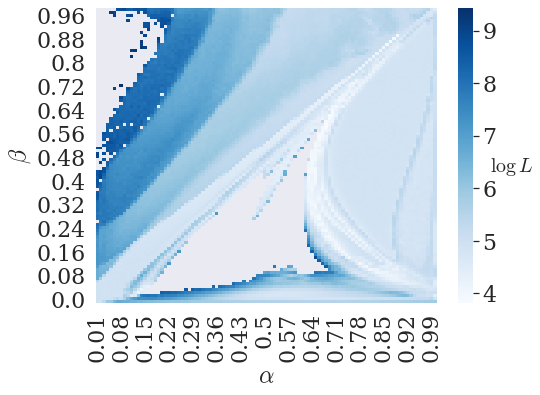}\par
        \includegraphics[width=\linewidth]{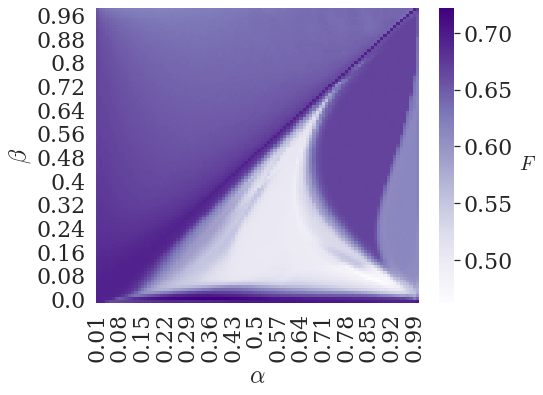}\par
    \end{multicols}
    \begin{multicols}{2}
        \includegraphics[width=\linewidth]{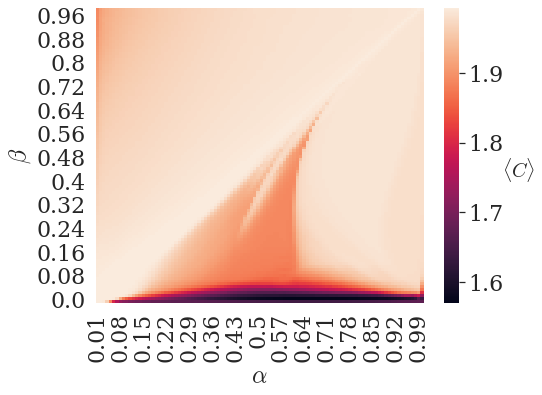}\par
        \includegraphics[width=\linewidth]{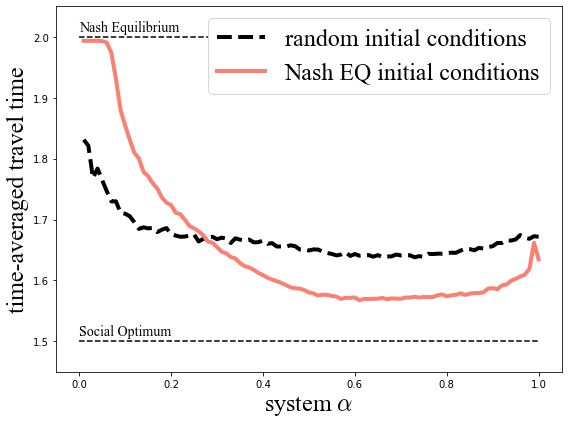}\par
    \end{multicols}
    \caption{(Row 1) $\beta=0$ for all experiments, left: period of cycles as a function of $\alpha$, right: probability of increase of cycles as a function of $\alpha$. (We explain the discontinuity near $alpha=0$ is likely due to measurement error \autoref{appendix:measurements}) (Row 2) full ablation study of $\alpha$ and $\beta$, left: $\log$ of the cycle period for color rendition, right: probability of increase of cycles. (Row 3) left: ablation study of $\alpha$ and $\beta$ for the time-averaged travel time, right: ablation of $\alpha$ setting $\beta=0$ for two initializations of $q$-values, random and Nash EQ.}
    \label{fig:cycle-study}
\end{figure}


\autoref{fig:cycle-study} shows our complete results for this study, completing the picture from the correlation matrices of \autoref{fig:correlation}. We see in \autoref{fig:cycle-study} how cycles are influenced by the learning rates, and how system performance is influenced by these cycles. They show clearly the overlap between the parameters that maximize system performance (minimize time-averaged travel time) for the same parameters that create the most "edgeworthy" (high $F$) and "fastest" (low $L$) cycles.


The plots of \autoref{fig:cycle-study} (Row 1) show the relationship between $\alpha$ and our cycle metrics, $L$ and $F$, for the case where $\beta=0$. $L$ tends to decrease as $\alpha$ increases, while $F$ increases, both reaching local minimum and maximum respectively for $\alpha \approx 0.6$. (Row 2) show these relationships while also varying $\beta$. The regions of interest are the cases where $\beta<\alpha$, so everything below the diagonal. Both plots feature a "sail-like" region below the diagonal, where both $L$ and $F$ are their lowest. 

(Row 3) plots the time-averaged travel times $\langle C \rangle$, on the left varying both $\alpha$ and $\beta$, and on the right varying only $\alpha$. The experiment is run for two different initializations of the $q$-values. The first $q$-value initialization is \textit{random}, where $q$-values are drawn uniformly at random from the interval $[1,2]$, the minimum and maximum travel costs. The second $q$-value initialization is \textit{Nash EQ}, where the $q$-values are set to $(2.1, 2.1, 2)$ for the actions (up, down, cross) respectively, for all agents. In this manner, at the start of the game all agents are induced to actions which result in the Nash equilibrium action profile. Both initializations yield a similar trend over $\alpha$ where small $\alpha$ values tend to yield higher travel times, and lower $\alpha$ values tend to decrease the travel time. Both initializations appear to have the best travel time performance for $\alpha \in [0.6, 0.8]$. As these specific parameter values are contingent on all other parameter settings, and hence somewhat arbitrary, we limit ourselves to claim that faster system $\alpha$'s tend to improve the time-averaged travel time, while slower system $\alpha$'s tend to worsen it.

The correlations of \autoref{fig:correlation} match the results of \autoref{fig:cycle-study}. The left correlation matrix shows the relationship between $\alpha,\beta$ and the dependent variables. It appears $\alpha$ only has a stronger negative correlation with $L$, while $\beta$ is positively correlated with $F$, $\langle C \rangle$ and negatively correlated with $\sigma_{\langle C \rangle}$. $\beta$ significantly worsens the performance of the system, and it seems $\alpha$ can not influence it much. However, if we set $\beta=0$, $\alpha$ is able to have a stronger influence on the system performance. The middle correlation matrix shows this relationship, and that $\alpha$ now has strong negative correlations with $L$ and $\langle C \rangle$, and stronger positive correlations with $F$ and $\sigma_{\langle C \rangle}$. The middle plot aligns with the results from \autoref{fig:cycle-study} (Row 1). Finally, the right correlation matrix shows the relationship between dependent variables also for the case where $\beta=0$. The asymmetry $F$ and periods $L$ are strongly negatively correlated. We can also see that $F$ is negatively correlated with $\langle C \rangle$, and conversely $L$ is positively correlated with $\sigma_{\langle C \rangle}$.

\section{Incentives and Learning Rates}\label{sec:incentives}

In the previous section we showed that the choice of the learning rate has a large influence on the period and asymmetry of the cycles that emerge in the repeated route-picking game played by continual $\epsilon$-greedy $Q$-learners. Importantly, we showed that cycles with short periods and high asymmetry coincide with the highest collusion. Therefore, we conclude that there are learning rates that can be picked to maximize collusion. However, is it strategically a best response for the $Q$-learning algorithm designers to pick these learning rates?

In this section we consider whether $Q$-learning algorithm designers which can strategically pick their parameters would end up in symmetric parameter assignments: playing the repeated route-picking game in a population of players all with the same parameters. To analyze this strategic parameter choice we define parameter-picking meta-games.

\paragraph{$x$-\textbf{Meta-Game}} 
Let $G=(n, (S_j)_{i=j \rightarrow |n|}, \langle C_j \rangle)$ be the strategic form game, where $n$ is the number of players, $S_j= [0,1]$ is the set of actions of player $j$ such that for $x_j\in S_j$, $x_j$ corresponds to a parameter for a $Q$-learner controlled by player $j$, and $\langle C_j \rangle$ is the time-averaged cost of player $j$. All players seek to minimize $\langle C_j \rangle$ by picking $x_j$. During a parameter-picking meta-game we assume that the other parameters are fixed and kept constant.

The action profile where every player picks the same action $x^*$ resulting in $\langle C^* \rangle$ for all players (i.e. all designers design their Q-learners in the same way) constitutes a \emph{symmetric pure-strategy Nash equilibrium} of the meta-game if, for every player $i$, there exists no $x_i\neq x*$ such that $\langle C_i \rangle$ given $i$ plays $x_i$ and all others play $x^*$ is lower than $\langle C^* \rangle$.

\paragraph{Best Response Evaluation} Next, we the set of symmetric pure-strategy action profiles for two parameter-picking meta-games: the $\alpha$-\textbf{Meta-Game} and the $\epsilon$-\textbf{Meta-Game}. The experimental setup follows that of \autoref{sec:cycles}, initializing $Q$-tables randomly. For analysis of the $\alpha$-\textbf{Meta-Game} we fix $\epsilon=0.01$. For analysis of the $\epsilon$-\textbf{Meta-Game} fix $\alpha=0.6$. To estimate the advantage that a player $j$ achieves by deviating to $x_j$, when all other players are evaluated for a fixed $\mathbf{x}_{-j}$ we define $D_j$, the difference between the time-averaged system cost $\langle C \rangle$ and the time-averaged cost for player $j$, $\langle C_j \rangle$:
\begin{equation}\label{eq:advantage}
    D_j = -(\langle C \rangle - \langle C_j \rangle).
\end{equation}
Note the negative operation since players seek to minimize their time-averaged travel time $\langle C_j \rangle$. In so doing we characterize unilateral best responses given the other players pick actions of a symmetric pure-strategy action profile.

\begin{figure}
    \begin{multicols}{2}
        \includegraphics[width=\linewidth]{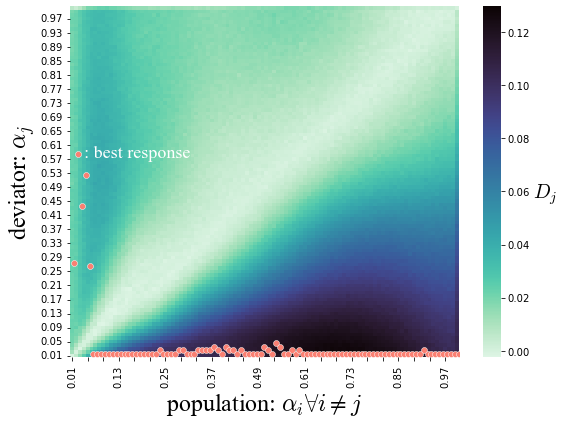}\par
        \includegraphics[width=\linewidth]{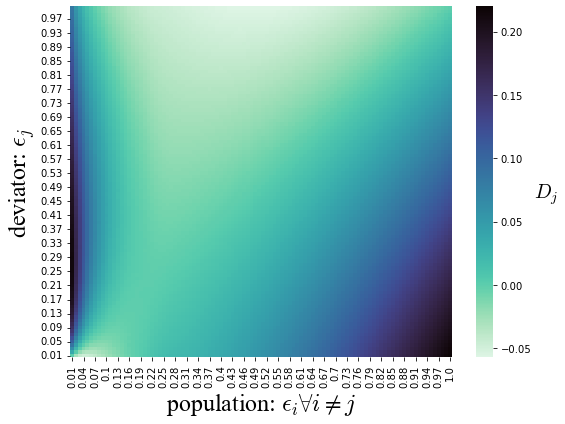}\par
    \end{multicols}
    \centering
    \caption{Advantage $D_j$
    (Eq.\ \ref{eq:advantage}) 
    of [left] an agent which picks alpha (vertical) against population of agents with fixed alpha (horizontal), [right] an agent which picks epsilon (vertical) against population of agents with fixed epsilon (horizontal)}
    \label{fig:alpha-heatmaps}
\end{figure}

Our key finding of the incentive compatibility of the meta-game of parameter picking summarizes as follows.

\paragraph{Result 1: There exists no symmetric pure-strategy Nash equilibrium} As \autoref{fig:alpha-heatmaps} shows, it is the case that, for a deviating player $j$, it is always better to deviate away from the parameter that the rest of the population picks. This is the case for both $\alpha$ [left plot] and $\epsilon$ [right plot]. This is visible from the clear diagonal in the left plot of \autoref{fig:alpha-heatmaps} (and a less clear but visible diagonal of the right plot) which always provides an advantage of around 0, and other parameter values for the deviating player yield greater advantage. From this result, two interesting corollaries follow.

\paragraph{Corollary 1: The social optimum is not incentive compatible} \autoref{fig:alpha-heatmaps} shows the best response $\alpha$ or $\epsilon$ given the rest of the population picks the same $\alpha$ or $\epsilon$. For the learning rate $\alpha$, from \autoref{sec:cycles} we know that the social optimal homogeneous $\alpha$ is about $0.6$, but we can clearly see that when all agents pick $\alpha=0.6$, the best response is to pick a very low $\alpha \approx 0.01$. For the exploration rate $\epsilon$, from literature \cite{carissimo2024counter} we know that $\epsilon=0.01$ is system optimal, but again we clearly see that when all agents pick $\epsilon=0.01$ the best response is to pick a larger $\epsilon$.

\paragraph{Corollary 2: Individual learners profit from being different} Low (high) parameters are best responses to a population with high (low) parameters. For example, we can see in \autoref{fig:alpha-heatmaps}, that it is always better to pick a low alpha when the $\alpha > 0.05$, and that when all players pick a low $\alpha$, which is a best response to high $\alpha$, then the best response becomes a high $\alpha$. A similar best response pattern applies for $\epsilon$.

\section{Discussion}\label{sec:discussion}

Our main results are that (i.) $Q$-learning, in particular continual $Q$-learning, 
can lead to tacit collusion in the form of cyclical phenomena akin to Edgeworth cycles, but also that (ii.) their emergence would require some form of explicit collusion in terms of parametrization. 


We shall use this final section to discuss these results for congestion games more narrowly, and to think about what these results might imply for other, similar game contexts like price setting games a la Bertrand.

\subsection{Coordination in Congestion Games}

The main difference of our approach and that of the previous literature is that we test $Q$-learning algorithms 
that undergo continual-learning. 
In this setup we find coordinated cyclical dynamics which resemble Edgeworth cycles,
that are stochastic and non-stationary.
Importantly, they 
are beneficial for the social welfare of the system. 
However, 
the individual incentives are not aligned, meaning that best responses in the $\alpha$-meta game may not lead to the best cycling behaviour; if 
the parameters of the $Q$-learning algorithms are 
being 
picked by myopic and self-interested algorithm designers,
which merely seek to minimize their travel costs. 
We also note that 
the individual incentives for picking exploration rates $\epsilon$ are not aligned, results which we include in \autoref{appendix:epsilon}.

To play the $\alpha$-meta game,
players must 
generate an internal representation for 
the expectation of the parameters of other players. If other players are expected to have low $\alpha$ parameters, then the best response is to pick a high $\alpha$. On the other hand, if other players are expected to have high $\alpha$ parameters (the socially optimal range) then the best response is to pick a low $\alpha$. It is not immediately clear how a player should best formulate this expectation, 
and our results strongly
point to the 
existence of a mixed equilibrium, characterized by 
a combination of both high and low $\alpha$. The existence of a mixed equilibrium, combined with the real limitation that only one $\alpha$ parameter can be picked at any time, may ultimately lead to a non-convergent dynamics in real systems where online $Q$-learning algorithms are applied by many independent players.

In this paper we considered the Braess Paradox, a notorious congestion game
whose 
features have been identified 
and studied 
in real-world systems \cite{fisk1981empirical, steinberg1983prevalence, arnott1994economics, tumer2000collective}. 
Examples include complex problems such as 
packet routing on the internet, 
and optimization in autonomous vehicle networks,
which 
may exhibit 
endogenous non-stationarities to which any algorithm must adapt \cite{park2005internet, yan2006efficient, helbing2001traffic}. Furthermore,
the internet and road networks 
are subsystems of 
large 
complex systems 
that may 
induce the emergence of 
exogenous non-stationarities 
that require suitable adaptation \cite{korecki2023deep, korecki2023well}. 
Our continual learning approach 
appears both feasible and suitable
for a wide range of real systems,
which feature congestion-like games.

Our results suggest, therefore, 
that (i) 
congestion games played by many $Q$-learners are unlikely to be found at a Nash 
equilibrium, and (ii) that the self-organized dynamics can yield better than average payoffs. 
In particular, 
a system designer with the power to set population-wide parameters should be able to choose near-optimal parameters -- cost-efficiently. 
Coordination, for example, 
requires
no additional communication or centralized control. 
In fact, the best performances are achieved when the individual learners only receive feedback for their chosen actions. 
Moreover, 
should players value (partially) the social welfare of the system, 
a system designer can benefit from 
simply 
recommending that all users pick a high value of $\alpha$.

Lastly, if agents have access to the feedback of others,
it is invariably 
a best response for 
them
to use it. We report these results in \autoref{appendix:beta-study}. However, universal access to this feedback has a detrimental effect for the self-organized coordination of the system. As such, if a system designer does not have 
full 
control of the parameters, as may be the case for a traffic control center, 
it is possible that 
reducing, or capping, 
the speed with which individuals can access the feedback of others 
would 
lead to 
strongly positive social welfare effects.


\subsection{Thoughts on Edgeworth cycles and collusion in pricing games}

The Braess Paradox shares one key characteristic with Bertrand games in that they both have Pareto inferior Nash Equilibria. Bertrand competition is where Edgeworth cycles have been shown to constitute equilibria, and where prior work on algorithms has shown that Q-learners can learn to collude tacitly in ways predicted by Edgeworth cycles. The main difference between the literature on $Q$-learning in Bertrand games and our approach is that we do not decay the exploration rate $\epsilon$ to study convergence. Instead, we study the learning dynamics of the $Q$-learners as they continually learn online while interacting in a repeated game. It is an open question to study whether Q-learning without decaying $\epsilon$ will also cycle in Bertrand games. Our analysis suggests that it would. Assuming that it would, this would open up several interesting questions.

In this paper, we were interested in reducing travel times in congestion games, which is why we ultimately aimed to enhance algorithmic coordination. By contrast, regulation is against algorithmic collusion in Bertrand competition, the opposite. Our results pose new policy challenges, if indeed the online continual learning approach that we have pursued here is also relevant for pricing games: in this case our results suggest that for the purpose of prevention of algorithmic collusion, collusion should not be seen as only occurring in the underlying Bertrand game, but also in the meta game of setting learning parameters. Our results show clearly that the best coordinated (collusive) behaviour is achieved for learning rates $\alpha$ (and exploration rates $\epsilon$ \autoref{appendix:epsilon}) which are not equilibria of the meta game. The best responses of players of the meta game do not align with one another, and it is always a best response to deviate from the population's average population parameters. Therefore, cycling behaviour in algorithmic pricing can be an indication that algorithm designers have colluded on setting the learning parameters of their algorithms, because Q-learning designed to maximize individual payoffs unilaterally would result in combinations of parameters that are not well suited for cycles to emerge. The presence of cycles with high average prices is therefore evidence of collusion at algorithm designer level, which might warrant development of a novel regulation perspective.

%
%
%
%
%

\bibliographystyle{ACM-Reference-Format}
\bibliography{references}

\newpage

\appendix

\section*{Appendices}

\section{Heterogeneous $\alpha$ Best Responses}

A complete analysis of the best responses of learning rates requires considering a heterogeneous population. We test this by fixing the mean of the population $\alpha$ at $0.5$, and then drawing the values for each agent from a uniform random distribution. We vary the variance of this distribution by enlarging the support starting from ${0.5}$ up to and including the full range $[0,1]$. We find that increasing the variance does not change the best response, but it does decrease the range of $\alpha$ values which yield "attractive" advantages. Therefore, we can conclude that in a population of $Q$-learners with heterogeneous $\alpha$ values, it may be harder to identify the best-response, but that the general principles still apply as found in \autoref{fig:alpha-heatmaps}.

\begin{figure}[h!]
        \includegraphics[width=0.5\linewidth]{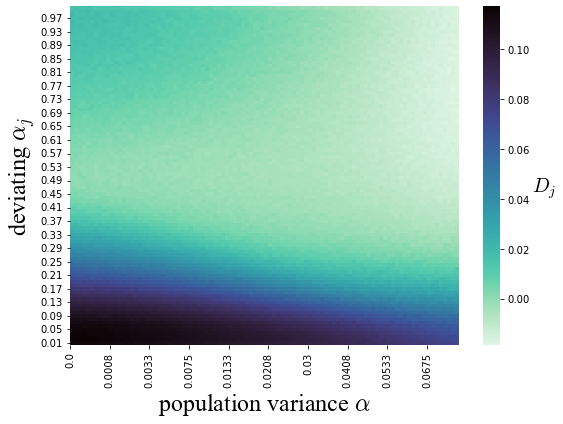}
    \centering
    \caption{
    Advantage $D_j$
    (Eq.\ \ref{eq:advantage}) 
    of an agent which picks alpha (vertical) against population of agents with alpha drawn from uniform distribution with mean $0.5$ and different variance (horizontal).}
    \label{fig:alpha-variance}
\end{figure}

\section{The Determinants of `Best' Parameters}\label{appendix:parameters}

\subsection{Alpha}

The determining factors of best responses for $\alpha$ are the \textit{stochasticity} and \textit{non-stationarity} of the system, as perceived from the perspective of a single agent.

In the left plot of Figure \ref{fig:alpha-heatmaps} we can see that when all other players pick $0.1 < \alpha < 1$ the best response for a deviating player is a very low $\alpha \approx 0.01$. This makes sense if we imagine that a high $\alpha$ from the other players leads them to update their $q$-values more, which will change their actions more, which will create a more stochastic environment. Then, the best strategy for a single player is to use a small $\alpha$ which will enable the player to average out more stochasticity and maintain an estimate of the true values of rewards which is more accurate. Then, let us look at the case where all other players pick a small $0<\alpha<0.05$, because we find here that the best response $\alpha$ for a deviating player is now large. When all other players have a small $\alpha$ their $q$-values will change slower, so their actions will change less and the game will become more stable and less stochastic. Thus, our deviating player benefits from picking an $\alpha$ which is higher (e.g. $\alpha=0.5$) which can converge faster to new observations and therefore exploit the non-stationarity of the system which undergoes cyclical behaviour.

\subsection{Beta}\label{appendix:beta-study}

\begin{figure}
    \begin{multicols}{2}
        \includegraphics[width=\linewidth]{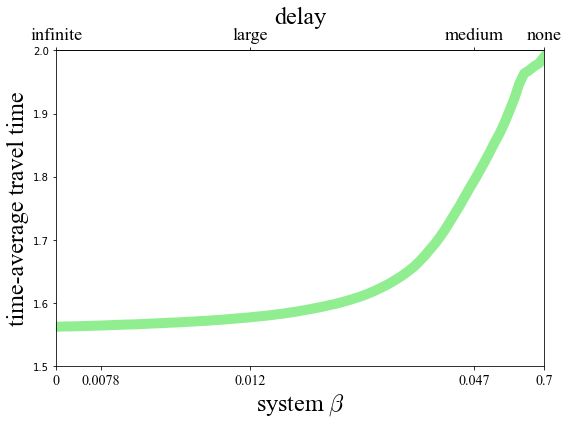}\par
        \includegraphics[width=\linewidth]{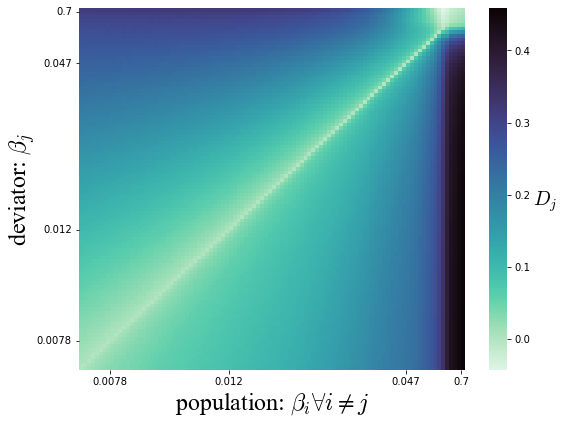}\par
    \end{multicols}
    \centering
    \caption{left: beta study considering that it is impossible to have 0 delays, so imitation always comes with a delay, right: best response $\beta$ given that no delays for imitation are impossible.}
    \label{fig:enter-label}
\end{figure}

Thus far we have considered an imperfect monitoring case where agents receive feedback as rewards only for the actions they have player. This monitoring creates delays in the $q$-values of agents which are then responsible for the cyclical dynamics we observe in the system. It is simple and natural to assume that agents have imperfect monitoring, because it is feasible to imagine that every agent can directly measure the cost of an action to use as feedback. Enhancing the monitoring abilities of an agents would require communication with other agents or with central planners. In this section we consider an extension of independent $Q$-learning where agents perceive the rewards for the other actions. In other words agents monitor the rewards for all actions.

We wish to understand the manner in which monitoring affects how an agent should best pick their learning rate. To do so we will introduce a new learning rate: the \textit{imitation} rate $\beta$. The imitation rate $\beta$ will be used to update the $q$-values for the actions that the agent did not pick. So, what value of $\beta$ should an agent choose? 

We explained when an agent should pick a low $\alpha$, high stochasticity, and when it should pick a high $\alpha$, low stochasticity and high non-stationarity. These principles remain similar for the imitation rate $\beta$ with a twist: we posit that the feedback we receive from the actions we played is immediate, but the feedback for imitation arrives with a delay. This delay may be due to an intrinsic delay in aggregation and communication, but it may just as well be an externally imposed delay by a central authority or regulating body. The extreme case where $\beta=0$ reflects a scenario where agents never get monitoring feedback. The other end of the extreme is where $\beta=\alpha$, and the monitoring feedback is immediate. In a case where monitoring feedback has a large delay, the feedback values will encode the past state of a game which may be less relevant to the present. In this case, we imagine that $\beta$ is small.

Like with the experiments from \autoref{sec:incentives} we will vary the imitation of the whole population and a deviating agent separately. In so doing, the imitation rate $\beta$ will always be kept smaller than the learning rate $\alpha$, $0 \leq \beta \leq \alpha \leq 1$.

What our results suggest is that any additional information that is provided to the agents about the results of other players can and will be used, and that players have full incentive to exploit this information. From the system designer perspective, or the perspective of a regulator which can enforce information sharing, no information should be shared. In fact, it may be beneficial to find ways to prevent information sharing.

\subsection{Epsilon}\label{appendix:epsilon}

We find very interesting results for the best responses of $\epsilon$ which we deemed too much for the main text of the paper. It appears that when all other players pick small values of $\epsilon$ that the best response $\epsilon$ is high, but that when all players pick high $\epsilon$, the best response $\epsilon$ is low. This is similar to the best response behaviour of $\alpha$, and can be explained in a similar manner. When all players pick low $\epsilon$ the system will have be stochastic. Picking a higher $\epsilon$ will then allow an agent to explore more, and realize quicker that "cross" is again a beneficial action to exploit. On the other hand, when all agents pick a high $\epsilon$ the system will be highly stochastic, and relatively stationary. This can be seen from the results in \cite{carissimo2024counter}. Therefore, an agent benefits from picking a low $\epsilon$ which does not explore much, as the underlying value of rewards will not be a moving target.

\begin{figure}
    \centering
    \begin{multicols}{2}
        \includegraphics[width=\linewidth]{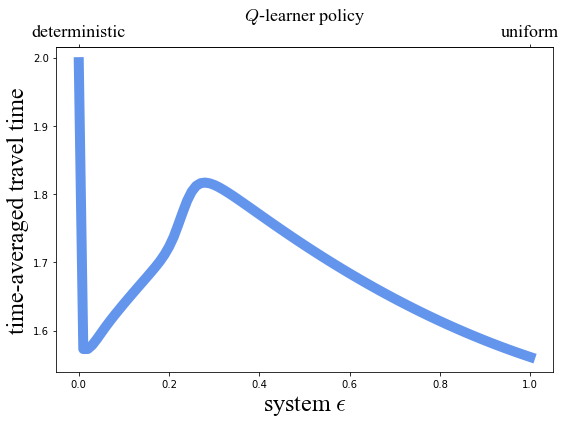}\par
        \includegraphics[width=\linewidth]{images/epsilon_study.png}\par
    \end{multicols}
    \caption{Left: $\langle C \rangle$ as a function of $\epsilon$, studied in greater detail in \cite{carissimo2024counter}. It appears low a values of $\epsilon$ can yield good system performance, due to implicit coordination of $Q$-learners. Right: The best response characterization of $\epsilon$, where a population is set to a fixed value, and a deviating agent evaluates the best response to the homogeneous population.}
    \label{fig:epsilon-study}
\end{figure}

\section{Measurements}\label{appendix:measurements}

\begin{figure}
    \centering
    \includegraphics[width=0.5\linewidth]{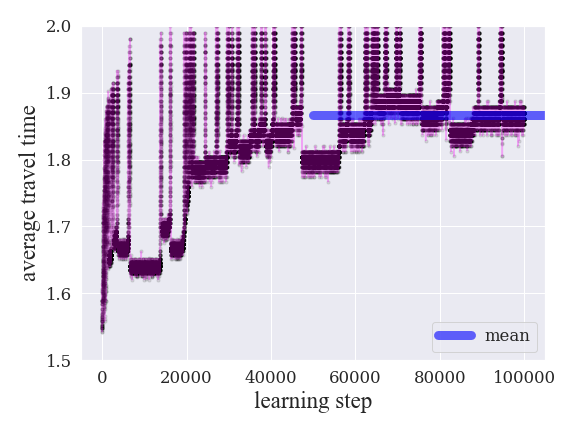}
    \caption{A single timeseries showing the mean for the second half of the series. The mean is far from the midpoint between the maximum and minimum of the timeseries, and is thus not a good threshold to evaluate the number of cycles in the time-window.}
    \label{fig:invalid-cycle}
\end{figure}

In \autoref{fig:cycle-study} (Row 1) we plot the measures of $L$ and $F$ as a function of $\alpha$, and find an odd kink in the measurements near $\alpha=0$. This is likely due to a measurement error of our method, rather than a true reflection of the period and asymmetry of the cycles. To explain this, we refer to \autoref{fig:invalid-cycle} where we show that the mean of the system, represented by the blue line, is no longer a suitable reference point to measure cycles. This is the case when $\alpha=0.01$, the lowest value of $\alpha$ we test, and appears to persist in \autoref{fig:cycle-study} for a few more values above $0.01$. Such a measurement error leads to measurements of low periods $L$, even though the actual period of the oscillations is much larger.

Similarly, in figure \autoref{fig:invalid-cycle} it is clear that the oscillations have a very low asymmetry, as they mostly proceeded flat. Nonetheless, the measured asymmetry as in \autoref{fig:cycle-study} (Row 1)(right) appears to high. We claim once again that this is an artefact of our measurement, rather than a true reflection of the behaviour that we intended to measure.

\end{document}